\documentclass{bmvc2k}
\usepackage{amsmath}
\usepackage{wrapfig,lipsum,booktabs}
\usepackage{mathtools}
\usepackage[outercaption]{sidecap} 
\usepackage{wrapfig,lipsum,booktabs}

\usepackage{times}
\usepackage{epsfig}
\usepackage{amsmath}
\usepackage{amssymb}
\usepackage{wrapfig,lipsum, booktabs}
\usepackage{pifont}
\usepackage[accsupp]{axessibility}  
\newcommand{\xmark}{\ding{55}}
\usepackage{multirow}
\usepackage{multicol}
\usepackage{color}
\usepackage{graphicx,multirow}
\usepackage{xcolor}

\title{MSA$^\text{2}$Net: Multi-scale Adaptive Attention-guided Network for Medical Image Segmentation}

\addauthor{Sina Ghorbani Kolahi}{sina.ghorbani@modares.ac.ir}{1}
\addauthor{Seyed Kamal Chaharsooghi}{skch@modares.ac.ir}{1}
\addauthor{Toktam Khatibi}{toktam.khatibi@modares.ac.ir}{1}
\addauthor{Afshin Bozorgpour}{afshin.bozorgpour@ur.de}{2}
\addauthor{Reza Azad }{reza.azad@rwth-aachen.de}{2,3}
\addauthor{Moein Heidari}{moein.heidari@ubc.ca}{4}
\addauthor{Ilker Hacihaliloglu}{ilker.hacihaliloglu@ubc.ca}{4}
\addauthor{Dorit Merhof }{dorit.merhof@ur.de}{2,5}
\addinstitution{
 Tarbiat Modares University\\
 Tehran, Iran
}
\addinstitution{
 University of Regensburg\\
 Regensburg, Germany
}
\addinstitution{
 RWTH Aachen University\\
 Aachen, Germany
}
\addinstitution{
 University of British Columbia\\
 British Columbia, Canada
}
\addinstitution{
Fraunhofer Institute for Digital Medicine MEVIS\\
Bremen, Germany
}
\runninghead{S. G. Kolahi, et al}{Multi-scale Adaptive Attention-guided Network}

\begin{document}

\maketitle

\begin{abstract}
Medical image segmentation involves identifying and separating object instances in a medical image to delineate various tissues and structures, a task complicated by the significant variations in size, shape, and density of these features. Convolutional neural networks (CNNs) have traditionally been used for this task but have limitations in capturing long-range dependencies. Transformers, equipped with self-attention mechanisms, aim to address this problem. However, in medical image segmentation it is beneficial to merge both local and global features to effectively integrate feature maps across various scales, capturing both detailed features and broader semantic elements for dealing with variations in structures. In this paper, we introduce MSA$^\text{2}$Net, a new deep segmentation framework featuring an expedient design of skip-connections. These connections facilitate feature fusion by dynamically weighting and combining coarse-grained encoder features with fine-grained decoder feature maps. Specifically, we propose a Multi-Scale Adaptive Spatial Attention Gate (MASAG), which dynamically adjusts the receptive field (Local and Global contextual information) to ensure that spatially relevant features are selectively highlighted while minimizing background distractions. Extensive evaluations involving dermatology, and radiological datasets demonstrate that our MSA$^\text{2}$Net outperforms state-of-the-art (SOTA) works or matches their performance. The source code is publicly available at \href{https://github.com/xmindflow/MSA-2Net}{\textit{https://github.com/xmindflow/MSA-2Net}}.

\end{abstract}

\section{Introduction}
\label{sec:intro}
Medical image segmentation is a crucial for disease diagnosis and quantitative assessment across the entire clinical workflow~\cite{antonelli2022medical}. This technology is especially critical in biomedical applications, where detecting individual tumors, organs, or cell entities and segmenting their shapes represents a major bottleneck~\cite{fiaz2022sa2}. The inherent challenges of automated segmentation arise from the considerable variations in the size, shape, appearance, and density of the object of interest. Deep learning networks excel in medical image analysis, outperforming traditional SOTA techniques. The fully convolutional neural network (FCN) stands out as one of the pioneering approaches employed for image segmentation~\cite{long2015fully}. The seminal U-Net~\cite{ronneberger2015u}, featuring a series of convolutional and down-sampling layers to gather contextual information along its contracting path, with skip connections from the encoder to up-sample both coarse and fine features, has demonstrated eminent segmentation potential.
The CNN and U-Net architectures have been enhanced through various adaptations aimed at optimizing their performance. Innovations have included the introduction of attention mechanisms~\cite{oktay2018attention}, the incorporation of dynamic modeling~\cite{shen2023dskca,li2019selective,azad2024beyond}, modifying skip connections~\cite{huang2020unet}, and the replacement of backbone modules within segmentation networks~\cite{karaali2022dr}. These enhancements have been critical in advancing the capabilities of these architectures to handle the complexities of medical image segmentation. While CNNs have traditionally been preferred for this task, they struggle to effectively model long-range dependencies and are limited by static receptive fields, which restrict their ability to capture the shape and structural details needed for precise medical image segmentation. The Vision Transformer (ViT)~\cite{dosovitskiy2020image} was designed to address this limitation. Equipped with the self-attention mechanism, ViTs excel at capturing global information and modeling long-range dependencies. However, transformers are limited in capturing the local representation and context and are computationally inefficient with quadratic complexity. Several methods have been proposed to overcome the disadvantages of the transformer model~\cite{Khan_2022}, such as hybrid CNN-Transformer architectures~\cite{heidari2023hiformer, sadegheih2024lhu}, reducing the computational cost of self-attention~\cite{shen2021efficient}, performing the self-attention mechanism in local regions~\cite{liu2021swin}, or calculating channel attention instead of spatial attention~\cite{ali2021xcit}. However, these methods capture only partially the global context while addressing the diverse challenges in medical images, such as variations in form, visual characteristics, and compactness, necessitating the integration of both local and global features.
A feasible approach in addressing the inherent limitations of the aforementioned architectures with respect to local and global information processing is the use of skip connections~\cite{azad2022medical}. In medical image segmentation, capturing both local details (like the texture of tissues) and global context (such as the structure of organs) is crucial. Skip connections help in preserving high-resolution details through the network layers, enabling the model to use both fine and coarse features for more precise segmentation. Skip connections assist CNNs by reintegrating precise local features from initial layers into deeper ones, blending detailed and broad information for comprehensive contextual understanding~\cite{dai2021attentional}, thus overcoming their naturally limited receptive fields. For transformers, skip connections enhance performance by integrating local details directly into layers focused on global processing, effectively balancing local precision with global awareness. However, most of these studies have focused solely on improving the encoder or decoder, and not on altering the original skip connection design as such~\cite{wang2022uctransnet}. Therefore, we propose MSA$^\text{2}$Net, augmented by innovative skip connections that incorporate our novel Multi-Scale Adaptive Spatial Attention Gate (MASAG). This module strategically highlights spatially relevant features across multiple scales, enabling it to effectively model complex structures for delineating and localizing objects of interest. This stage facilitates a spatial interplay between local and global features, enriching the feature maps with a nuanced, contextually informed representation~\cite{fan2024lightweight}. Within the decoder, initial layers employ the Large Kernel Attention (LKA) module~\cite{guo2023visual} for adept local and global feature management, addressing typical high-resolution image processing challenges in transformers. Deeper layers of the decoder use Dual Attention Enhanced Transformer blocks (DAE-Former)~\cite{azad2023dae}, effectively preserving long-range interactions in lower-resolution images while incorporating both spatial and channel attentions seamlessly, thus maintaining the grid integrity of the images. This strategy guarantees accurate and context-sensitive delineation of anatomical features within intricate medical images.

\noindent\textbf{Contributions:} The key contributions of this paper are as follows:
(i) The introduction of MASAG, which dynamically recalibrate the receptive fields to emphasize spatially relevant features and suppress irrelevant background details. (ii) A hybrid decoder that integrates DAE-Former blocks in deeper layers for low-resolution image processing and LKA modules in shallow layers for high-resolution detail management for accurate and boundary-aware segmentation. (iii) Extensive evaluation across two challenging medical datasets, namely dermatology (ISIC2018~\cite{codella2019skin}), and radiology (Synapse~\cite{synapse2015ct}) datasets, demonstrates that our method surpasses SOTA approaches across diverse metrics.
\section{Related Work}
\label{sec:related}
\textbf{CNN-based Segmentation Networks:}
Motivated by the remarkable success of the U-Net model~\cite{ronneberger2015u}, it has been widely adopted for various medical image segmentation tasks through backbone, bottleneck or skip connection enhancements~\cite{azad2022medical, huang2020unet}.
Oktay et al.~\cite{oktay2018attention} introduced a novel attention gate (AG) module to emphasize key features relevant to a specific task. Shen et al.~\cite{shen2023dskca} added a dynamic selective attention module based on selective convolution kernel~\cite{li2019selective} for learning the dependencies between the channels. However, in our method, channel attention has been altered to spatial attention to suppress irrelevant areas and highlight salient regions based on dynamic receptive fields to learn local and global contextual information.

\noindent\textbf{Transformer-based and Hybrid Segmentation Networks:}
Transformers have demonstrated their capability to gather global information, addressing the drawback of CNNs in capturing long-range semantic dependencies and limited receptive fields.
Some of these studies aim to capture long-range spatial context using purely transformer-based approaches instead of CNN-based methods, which partially broaden the receptive field of CNNs~\cite{valanarasu2021medical}. However, the quadratic computational complexity of transformers, along with their inefficiency in capturing local contextual relationships between intra-path pixels, limits their usability in dense vision tasks such as medical image segmentation, which requires the neighboring pixel dependencies to be considered in the multi-scale and hierarchical pattern~\cite{khan2023recent}. To concurrently exploit the multi-scale representation along with the ability to capture local semantic and texture information of CNNs, and long-range interaction of transformers, cohort studies are dedicated to hybrid CNN-transformer based methods~\cite{chen2021transunet, sadegheih2024lhu, azad2024beyond}.
Specifically, Swin-UNet~\cite{cao2022swin} and DS-TransUNet~\cite{lin2022ds} introduce models with a U-shaped structure that utilizes the Swin Transformer~\cite{liu2021swin} for 2D image segmentation.

\noindent\textbf{Improving Skip-connections in Segmentation Networks:} Methods for redesigning skip-connections broadly attempt to process the encoder feature maps passed through the skip connections~\cite{xiang2020bio}, combine the
two sets of feature maps~\cite{oktay2018attention}, or extend the number of skip-connections~\cite{li2019cr}. Despite these efforts, previous methods often encounter limitations with fixed receptive fields and simplistic fusion methods in skip connections, such as addition or concatenation~\cite{Rahman_2023_WACV, oktay2018attention, ronneberger2015u}, which overlook content variance present in medical images.
To address this, we incorporate MASAG into our hybrid CNN-Transformer, which provides a multi-scale strategy and enhanced feature fusion. MASAG dynamically tunes spatial receptive fields and applies selective spatial weights, sharpening the focus on essential features while minimizing background noise and reducing false positives. This approach not only enriches the feature maps with local-global contextual information, crucial for accurate segmentation of varied organ shapes and sizes, but also recalibrates the encoder's output with a detailed spatial attention map.

\section{Methodology}
\label{sec:method}
\subsection{Motivation}\label{method:subsec:motivation}
Our methodology, motivated by the complexity of medical image segmentation and the limitations of previous architectures, employs a context-sensitive framework with adaptive skip connections to better capture anatomical details. This adaptive framework is pivotal in merging the encoder's high spatial resolution with the decoder's semantically rich features for optimized feature fusion. Essential to this process is a module within the skip connections that dynamically accentuates vital features and selectively focuses on relevant regions, harmonizing local details with a broader contextual understanding. This strategy is critical to minimizing background noise and highlighting key areas, underpinned by a boundary-aware approach in both the proposed module and the loss function. Furthermore, the deliberate configuration of decoder networks is key, ensuring the cohesive fusion of detailed and semantic features, while addressing the local and global feature processing gaps inherent in transformer and convolutional models.

\subsection{Overall Architecture}\label{method:subsec:arch}
\begin{figure}
    \centering
    \includegraphics[width=\textwidth]{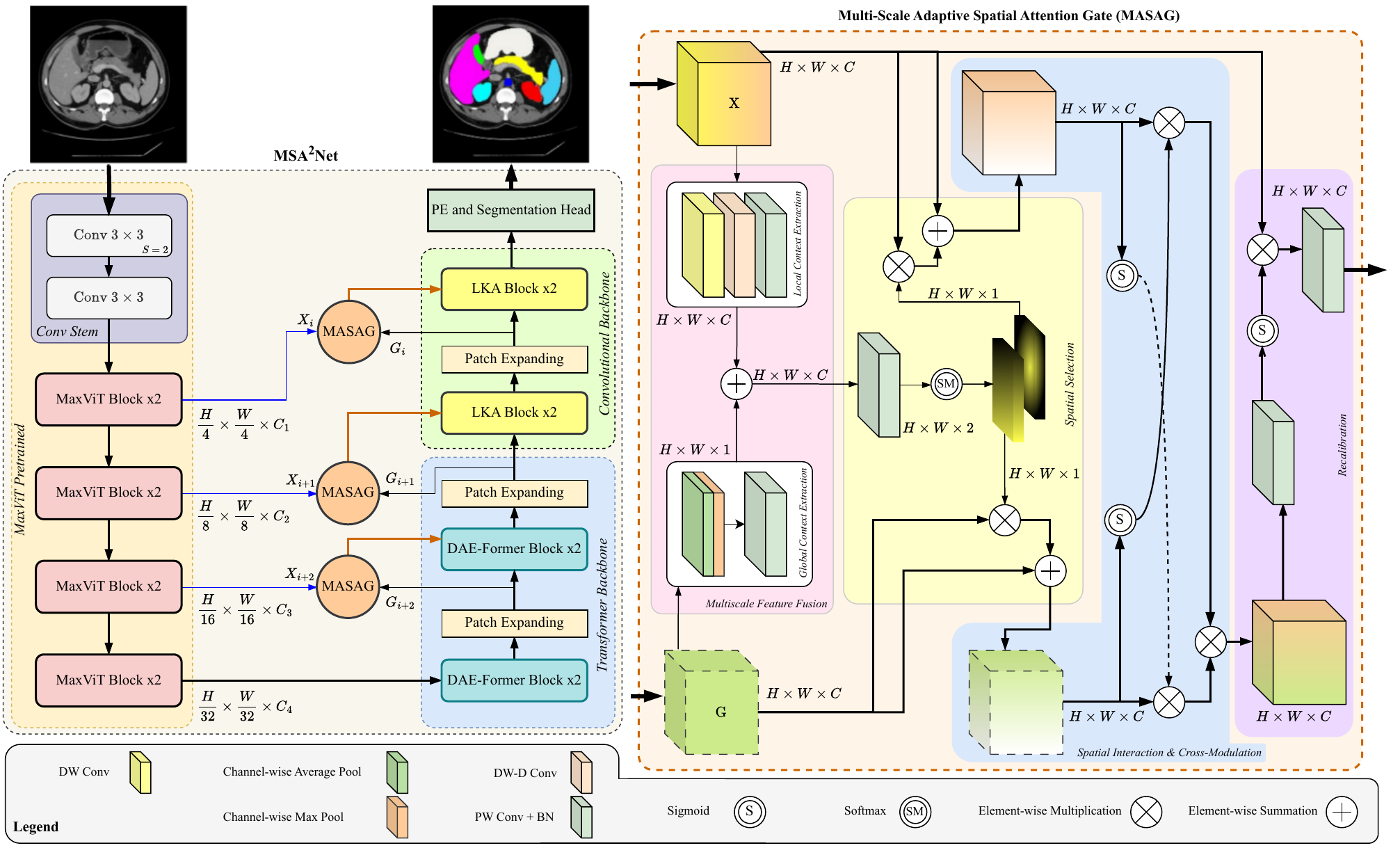}
    \caption{Our proposed segmentation network, called MSA$^\text{2}$Net, is composed of an encoder (using pretrained \textit{MaxViT} block) and a decoder (comprising DAE-Former blocks in deeper layers and LKA blocks in shallow ones). The encoding-decoding feature fusion is performed via our novel MASAG module.}
    \label{fig:msa2net}
\end{figure}

The overall architecture of our novel method, named MSA$^\text{2}$Net, as illustrated in Figure \ref{fig:msa2net}.
The proposed method leverages the MaxViT~\cite{tu2022maxvit} hierarchical transformer in the encoder to effectively capture multi-scale features through a combination of MBConv~\cite{sandler2018mobilenetv2} and various attention mechanisms. Enhanced by innovative skip connections that include our novel MASAG module, the method ensures that spatially relevant features are selectively emphasized while background distractions are minimized due to dynamically adjust of the spatial receptive fields, guided by inputs from both the encoder and decoder. Moreover, MASAG introduces a spatial interaction stage, enabling a bidirectional process between local and global features, thus providing the feature maps with a detailed, contextually informed representation. The decoder's shallow layers employ the LKA~\cite{guo2023visual} modules to adeptly balance local and global features as well as channel interactions, overcoming typical high-resolution image processing challenges associated with self-attention. Deeper decoder layers utilize DAE-Former blocks~\cite{azad2023dae} to efficiently maintain long-range dependencies in lower-resolution images, integrating spatial and channel attentions without disrupting the 2D structure of images. Overall, this method aims at accurate, context-aware segmentation of anatomical structures in complex medical images.

\subsection{Multi-Scale Adaptive Spatial Attention Gate (MASAG)}\label{method:subsec:adaptive}
Addressing the variability in organ structures, sizes, and shapes is a significant challenge. Traditional methods often rely on fixed receptive fields and simplistic fusion techniques in skip connections, failing to capture content diversity. To address this, we introduce the MASAG within our hybrid CNN-Transformer architecture's skip connections to enhance segmentation accuracy. MASAG, with its four stages -- \textit{multi-scale fusion}, \textit{spatial selection}, \textit{spatial interaction and cross-modulation}, and \textit{recalibration} -- processes encoder and decoder outputs (\( \mathbf{X} \) and \( \mathbf{G} \), respectively). The multi-scale fusion is utilized to aggregate \( \mathbf{X} \) and \( \mathbf{G} \) in a semantically similar way, preparing the fused feature map for subsequent processing. Spatial selection dynamically adjusts receptive fields to prioritize essential features, while spatial interaction and cross-modulation enrich feature maps with both local details and global context. In the recalibration stage, the initial input from encoder is refined using a dynamically generated, selective attention map from MASAG, ensuring spatial precision in segmentation. This approach allows MASAG to effectively navigate the diverse challenges of medical imaging, guaranteeing precise and dependable segmentation outcomes. Each stage of MASAG will be explained in detail, elucidating the structured approach and operational workflow, as highlighted in Figure \ref{fig:msa2net} (right side).

\subsubsection{Multi-Scale Feature Fusion}\label{method:subsubsec:fusion}
The MASAG module's multi-scale feature fusion level integrates \textit{Local Context Extraction} and \textit{Global Context Extraction} to blend the encoder's high-resolution spatial details $\left(\mathbf{X}\right)$ with the decoder's semantic information $\left(\mathbf{G}\right)$. \textit{Local Context Extraction} broadens the spatial scope of $\mathbf{X}$ using depthwise and dilated convolutions, while \textit{Global Context Extraction} captures wide-ranging contextual information from $\mathbf{G}$ via channel-wise poolings. This integration forms a comprehensive feature map, crucial for the Spatial Selection stage in MASAG, ensuring accurate segmentation across diverse anatomical features. The multi-scale feature fusion is formulated as follows:
\begin{flalign}
\begin{aligned}
\label{eq1}
\mathbf{U} = \underbrace{\text{Conv}_{1 \times 1 }(\text{DW-D}\left(\text{DW}\left(\mathbf{X}\right)\right))}_{\textit{Local Context Extraction}} + \overbrace{\text{Conv}_{1 \times 1}\left(\left[\text{P}_{Avg}(\mathbf{G}); \text{P}_{Max}(\mathbf{G})\right]\right)}^{\textit{Global Context Extraction}}.
\end{aligned}
\end{flalign}\\
\noindent Here, DW and DW-D denote depthwise and dilated convolutions, and \(\text{Conv}_{1 \times 1}\) is point-wise convolution for \textit{Local Context Extraction}, while \(\text{P}_{\text{Max}}\) and \(\text{P}_{\text{Avg}}\) represent max and average pooling for \textit{Global Context Extraction}, The notation [;] signifies concatenation.

\subsubsection{Spatial Selection}\label{method:subsubsec:selection}

In the Spatial Selection stage of MASAG, the fused feature map $\left(\mathbf{U}\right)$, obtained from Equation~\ref{eq1} is projected to two channels, aligning with inputs $\mathbf{X}$ and $\mathbf{G}$. Spatially selective weights, calculated via a channel-wise softmax, refine attention across $\mathbf{X}$ and $\mathbf{G}$, leading to spatially selected versions $\mathbf{X}^\prime$and $\mathbf{G}^\prime$. These versions emphasize key areas while minimizing less relevant ones, enhancing receptive field selection. Additionally, two residual connections for $\mathbf{X}^{\prime}$ and $\mathbf{G}^{\prime}$ improve gradient flow and feature utilization, further refining the segmentation with precise, context-aware attention. The formulation of this process is as follows:
\begin{gather}  
\text{SW}_{i} = S\left(\text{Conv}_{1\times1}(\mathbf{U})\right) \quad \forall i \in [1,2], \label{eq2}\\ 
\mathbf{X}^{\prime} = \text{SW}_{1} \otimes \mathbf{X} + \mathbf{X}, \quad \mathbf{G}^{\prime} = \,\text{SW}_{2} \otimes \mathbf{G} + \mathbf{G}, \label{eq3}    
\end{gather}
where $\text{SW}_{i} \in \mathbb{R}^{2 \times \mathrm{H} \times \mathrm{W}}$, derived via channel-wise softmax $(S)$, ensures that the weights across two channels $(i \in [1,2])$ sum to $1$ at each spatial location, indicating the relative  contribution of inputs according to content.
\subsubsection{Spatial interaction and Cross-Modulation}\label{method:subsubsec:inter}
The Spatial Interaction and Cross-Modulation level within MASAG \textit{dynamically} enhances segmentation by weaving together the spatially refined inputs 
$\mathbf{X}^\prime$ and $\mathbf{G}^\prime$ (derived from Equation~\ref{eq3}). Here, $\mathbf{X}^\prime$ with its refined local detail, is enriched by global contexts from $\mathbf{G}^\prime$, through spatial weights applied via a \textit{sigmoid}, creating $\mathbf{X}^{''}$. Conversely, $\mathbf{G}^\prime$, initially embodying weighted broader contexts, incorporates detailed contexts from $\mathbf{X}^\prime$, evolving into $\mathbf{G}^{''}$. This mutual enhancement ensures that both 
$\mathbf{X}^{''}$ and $\mathbf{G}^{''}$ integrate a comprehensive mix of local and global contexts, solidified by fusing $\mathbf{X}^{''}$ and $\mathbf{G}^{''}$ via multiplication,
this integration combines detailed precision with overarching insights, effectively segmenting complex structural variations.
The formulation of spatial interaction is described as follows:
\begin{gather}
\label{eq4}
    \mathbf{X}^{''} = \mathbf{X}^{\prime} \otimes \sigma \left(\mathbf{G}^{\prime}\right), \quad \mathbf{G}^{''} = \mathbf{G}^{\prime} \otimes \sigma \left(\mathbf{X}^{\prime}\right),\\
    \mathbf{U}^{\prime} = \mathbf{X}^{''} \otimes \mathbf{G}^{''},\label{eq5}
\end{gather}
where, $\sigma \left(\mathbf{G}^{\prime}\right)$ and $\sigma \left(\mathbf{X}^{\prime}\right)$ refer to local and global spatial weights respectively, and $\mathbf{U}^{'}$ denotes multiplied fused feature map.

\subsubsection{Recalibration}\label{method:subsubsec:recal}

In the Recalibration stage, the fused feature map from Spatial Interaction and Cross-Modula-\newline tion (Equation~\ref{eq5}) is refined through a point-wise convolution, followed by a \textit{sigmoid} activation, to create a focused attention map. This map is used to recalibrate $\mathbf{X}$, the initial input from the encoder, by multiplying it with the attention map and further processing through another point-wise convolution layer, enhancing $\mathbf{X}$ with an adaptive multi-scale receptive field tailored by previous stages. This ensures $\mathbf{X}$ is aptly prepared with precise, context-aware features for integration into the decoder, facilitating accurate segmentation. The detailed formulation of this recalibration process is shown as follows:
\begin{align}
    \mathbf{X} = \text{Conv}_{1 \times 1}\left(\sigma\left(\text{Conv}_{1 \times 1}\left(\mathbf{U}^{\prime}\right)\right) \otimes \mathbf{X}\right).
\end{align}

\section{Experiments}
\subsection{Datasets}
To evaluate our method, we use the Synapse dataset~\cite{synapse2015ct} that encompasses a multi-organ segmentation across 30 cases with 3779 axial abdominal clinical CT images. We align our evaluation protocol with the guidelines provided by~\cite{chen2021transunet}. Further, we extend our evaluation to the skin lesion segmentation challenge, using the ISIC 2018 dataset~\cite{codella2019skin}, which includes 2594 dermoscopy images paired with their respective ground truth annotations. We adopt the data splitting and preprocessing procedures as outlined in~\cite{asadi2020multi}. 

\subsection{Implementaion Details}
Our Model, developed using the PyTorch library, operates on an NVIDIA RTX 3090 GPU and processes input images resized to 224 x 224 pixels. We used the pretrained MaxViT model for encoding. For the Synapse dataset, we configured a batch size of 20, an SGD solver with a base learning rate of 0.05, a momentum of 0.9, and a weight decay of 0.0001, training over 700 epochs. The ISIC dataset employed a batch size of 16 and was trained for 50 epochs using an Adam optimizer with an initial learning rate of 0.0001. Additionally, our model incorporates the Boundary Difference over Union (BDoU) Loss~\cite{sun2023boundary} to enhance boundary-aware segmentation, dynamically adjusting penalties based on object size.
\subsection{Quantitative Results}
Tables \ref{tab:synapse} and \ref{tab:isic} provide a comprehensive comparison of our proposed MSA$^\text{2}$Net method with previous SOTA approaches. In the Synapse dataset (Table \ref{tab:synapse}), MSA$^\text{2}$Net shows superior performance over the 2D version of D-LKA by 0.48\%. D-LKA is a hybrid model that employs deformable convolution in its LKA block to dynamically adjust the receptive field, effectively capturing organs in various shapes, sizes, and structures. Our method also surpasses DAE-Former, which relies on a fully transformer architecture akin to MISSFormer and Swin-Unet, by 2.21\%. Additionally, MSA$^\text{2}$Net outperforms Hiformer, a multi-scale CNN-Transformer model, by 4.36\% in Dice Similarity Coefficient (DSC) and achieves a 1.41 reduction in Hausdorff Distance (HD95) compared to the same.
Our method exhibits notable improvements in the segmentation of the spleen, aorta, liver, pancreas, and left kidney, with particularly significant enhancements in the pancreas and aorta segments, where increases of 1.59\% and 1.13\% are observed over the second-best method, respectively. The quantitative results from the Synapse dataset illustrate that models utilizing dynamic receptive field recalibration yield better outcomes for both small and large organs, showcasing MSA$^\text{2}$Net's advantages over 2D D-LKA.
For the ISIC dataset, as detailed in Table \ref{tab:isic}, our model not only achieves the highest scores across most metrics, but also surpasses other methods by at least 1.24\% in DSC. This improvement can be attributed to the effectiveness of our multi-scale feature fusion modules in better capturing features of the boundary areas. These collective findings underscore the efficacy of dynamic receptive field adjustments in our MSA$^\text{2}$Net approach, confirming its potential to set new benchmarks in the field of medical image segmentation.

\begin{table}[!ht]
    \centering
    \resizebox{\textwidth}{!}{\begin{tabular}{l||c|c||cccccccc||cc}
    \toprule
     \multirow{2}{*}{Methods}& \multirow{2}{*}{Params (M)}& \multirow{2}{*}{FLOPs (G)}& \multirow{2}{*}{Spl.}&  \multirow{2}{*}{RKid.}& \multirow{2}{*}{LKid.}&  \multirow{2}{*}{Gal.}&  \multirow{2}{*}{Liv.}&  \multirow{2}{*}{Sto.}& \multirow{2}{*}{Aor.}& \multirow{2}{*}{Pan.}& \multicolumn{2}{c}{Average}\\
     \cline{12-13}
     & & & & & & & & & & &DSC$\uparrow$ &HD95$\downarrow$\\

     \midrule\midrule
     
     TransUNet~\cite{chen2021transunet}& 96.07& 88.91&85.08 &77.02 &81.87 &63.16 &94.08 &75.62 &87.23  &55.86 &77.49 &31.69\\
     
     Swin-UNet~\cite{cao2022swin}& 27.17& 6.16& 90.66& 79.61& 83.28& 66.53& 94.29& 76.60& 85.47& 56.58& 79.13& 21.55\\
     
     
     MISSFormer~\cite{huang2021missformer}& 42.46& 9.89& \textcolor{red}{91.92}& 82.00& 85.21& 68.65& 94.41& 80.81& 86.99& 65.67& 81.96& 18.20\\
     
     ScaleFormer~\cite{huang2022scaleformer}& 111.6& 48.93& 89.40& 83.31& 86.36& \textcolor{blue}{74.97}& \textcolor{red}{95.12}& 80.14& 88.73& 64.85& 82.86& 16.81\\
     
     HiFormer-B~\cite{heidari2023hiformer}& 25.51&  8.045& 90.99& 79.77&  85.23& 65.23& 94.61& 81.08& 86.21& 59.52& 80.39& \textcolor{red}{14.70}\\
     
     DAEFormer~\cite{azad2023dae}& 48.07& 27.89& 91.82& 82.39& 87.66& 71.65& 95.08& 80.77& 87.84& 63.93& 82.63& 16.39\\
     
     PVT-CASCADE~\cite{Rahman_2023_WACV}& 35.28& 6.40& 90.10& 80.37& 82.23& 70.59& 94.08& 83.69& 83.01& 64.43& 81.06& 20.23\\
     
     
     2D D-LKA Net~\cite{azad2024beyond}& 101.64& 19.92& 91.22& \textcolor{blue}{84.92}& \textcolor{blue}{88.38}& 73.79& 94.88& \textcolor{blue}{84.94}& \textcolor{red}{88.34}& \textcolor{red}{67.71}& \textcolor{red}{84.27}& 20.04\\

     \midrule
     
     \textbf{MSA$^{\text{2}}$Net (Ours)}& 112.77& 15.56& \textcolor{blue}{92.69}& \textcolor{red}{84.24}& \textcolor{red}{88.30}& \textcolor{red}{74.35}& \textcolor{blue}{95.59}& \textcolor{red}{84.03}& \textcolor{blue}{89.47}& \textcolor{blue}{69.30}& \textcolor{blue}{84.75}& \textcolor{blue}{13.29}\\
     \bottomrule
      \end{tabular}}\vspace{0.5em}
    \caption{The evaluation outcomes of the suggested approach on the Synapse dataset are presented in this comparison. The highest performing result is highlighted in \textcolor{blue}{blue}, while the second highest is indicated in \textcolor{red}{red}. The metrics include parameters, reported in millions (M), and FLOPS, quantified in billions (G). The DSC and HD95 values are provided for several abdominal organs: spleen (Spl), right kidney (RKid), left kidney (LKid), gallbladder (Gal), liver (Liv), stomach (Sto), aorta (Aor), and pancreas (Pan).}
    \label{tab:synapse}
\end{table}

\begin{table}[!tbh]
    \centering
    \scalebox{0.7}{
        \begin{tabular}{l || c c c c }
            \hline
            \textbf{Methods}                    & \textbf{DSC}    & \textbf{SE}     & \textbf{SP}     & \textbf{ACC}    \\
            \hline
            U-Net \cite{ronneberger2015u}       & 0.8545          & 0.8800          & 0.9697          & 0.9404          \\
            AttU-Net \cite{oktay2018attention} & 0.8566          & 0.8674          & \textcolor{blue}{0.9863} & 0.9376          \\
            TransUNet \cite{chen2021transunet}  & 0.8499          & 0.8578          & 0.9653          & 0.9452          \\
            FAT-Net \cite{fatnet}          & 0.8903          & \textcolor{blue}{0.9100} & 0.9699          & 0.9578          \\
            Swin-UNet \cite{cao2022swin}      & 0.8946          & \textcolor{red}{0.9056 }         & 0.9798          & \textcolor{blue}{0.9645}  \\
            UCTransNet \cite{wang2022uctransnet} & 0.8838 & 0.8429 & \textcolor{red}{0.9825} & 0.9527 \\
            
            DermoSegDiff~\cite{bozorgpour2023dermosegdiff} & \textcolor{red}{0.9005} & 0.8761 & 0.9811 & 0.9587 \\

            \hline
            MSA$^{\text{2}}$Net (Ours)  & \textcolor{blue}{0.9129} & 0.8840 & 0.9557 & \textcolor{red}{0.9640}  \\
             \hline
        \end{tabular}
    }
    \caption{Comparative analysis of the proposed method versus SOTA techniques on ISIC2018 benchmark. \textcolor{blue}{Blue} indicates the best result, and \textcolor{red}{red} displays the second-best.}
    \label{tab:isic}
\end{table}

\subsection{Qualitative Results}
\begin{figure*}[!ht]
    \includegraphics[width=\textwidth]{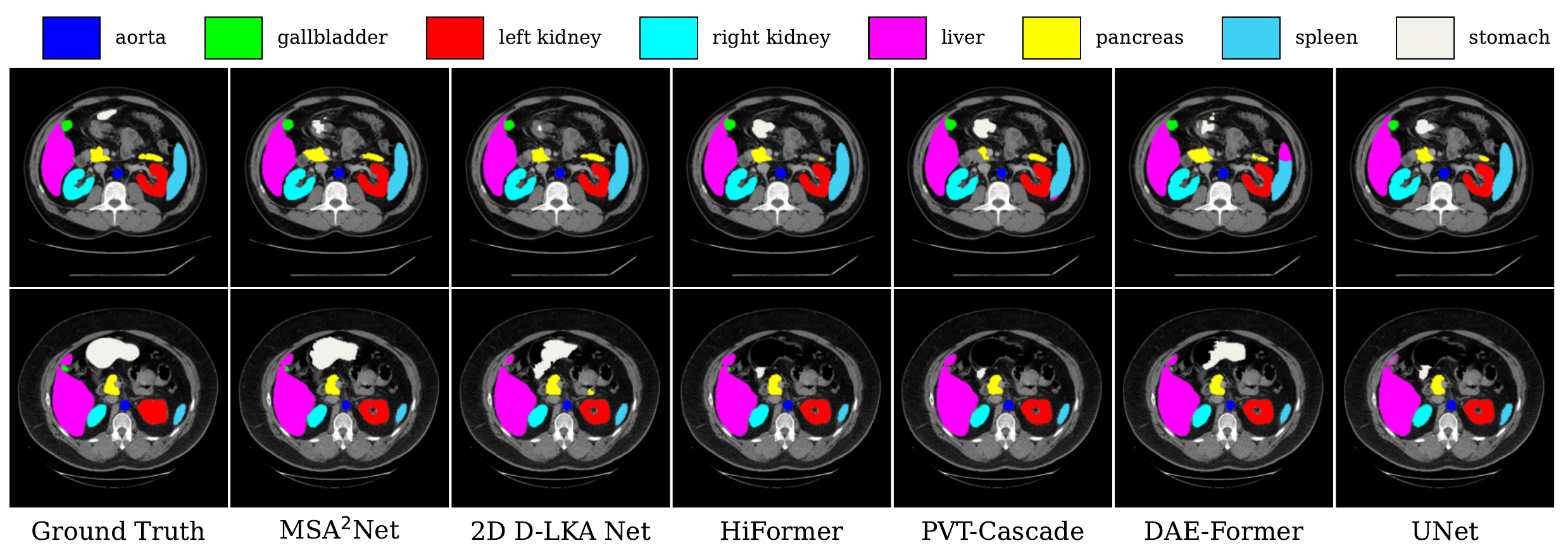}
    \caption{A comparative visual examination of the proposed approach in contrast to different methods employing the Synapse multi-organ segmentation dataset.}
    \label{fig:er-synapse}
\end{figure*}

\begin{figure*}[!ht]
    \centering
    \includegraphics[width=0.55\textwidth]{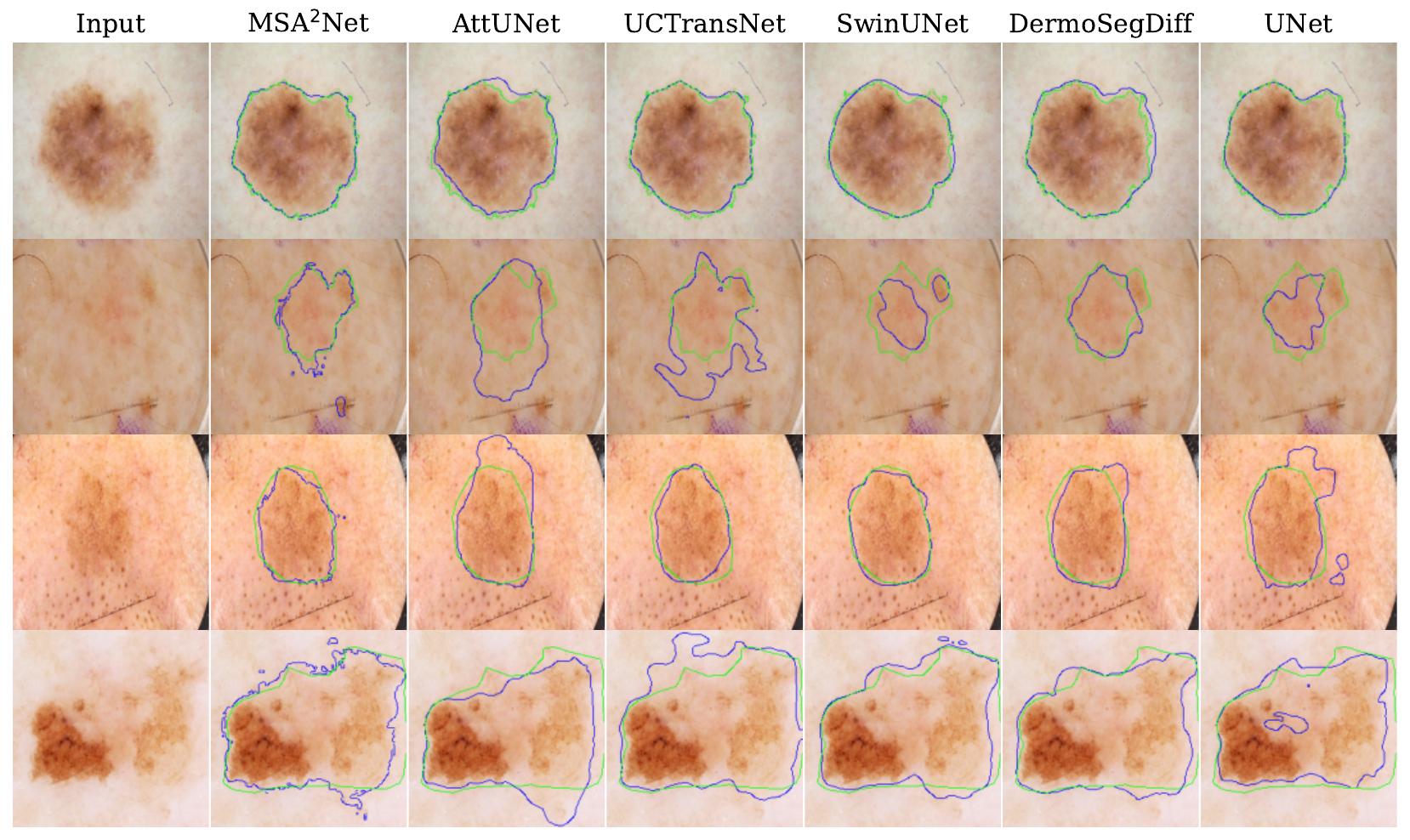}
    \caption{Visual comparisons of various techniques on the \textit{ISIC2018} skin lesion segmentation dataset are depicted. Authentic boundaries are represented in \textcolor{green}{green}, while anticipated boundaries are depicted in \textcolor{blue}{blue}.}
    \vspace{-1.5em}
    \label{fig:er-isic2018}
\end{figure*}

We showcase visual segmentation results for the Synapse (Figure \ref{fig:er-synapse}) and ISIC (Figure \ref{fig:er-isic2018}) datasets to further demonstrate the capacity of our method. It is evident that the predictions closely match the actual anatomical structure of the organs in the Synapse dataset, maintaining detailed features at the edges and precisely identifying organ areas. Furthermore, a visual assessment of the ISIC dataset shows that the boundaries predicted by our proposed model closely align with the ground truth, surpassing earlier methods, particularly in complex cases characterized by intricate color distribution and irregular borders, owing to our boundary-aware model and the MASAG module, which ensures optimal feature fusion.

\subsection{Ablation Study}
We conducted an ablation study on the Synapse dataset to examine the effectiveness of our modules under various configurations as summarized in Table \ref{table:ablation}. 
Specifically, the inclusion of MASAG (last three rows) demonstrates its capability to fine-tune focus adaptively, emphasizing key areas and dynamically adjusting features to enhance segmentation, as evidenced by improved DSC and reduced HD scores. Besides, the final row demonstrates that our design elements work together effectively, enhancing segmentation predictions by leveraging the synergistic effects of all components for superior detail preservation and precise organ delineation.
\begin{table}[!hbt]
    \centering
        \scalebox{0.7}{
    \begin{tabular}{cccccccc}
    \hline
     \textbf{Model} & \textbf{MASAG} & \textbf{LKA}  & \textbf{DAE-Former} & \textbf{Dice} & \textbf{HD95}
     \\ \hline
     MSA$^{\text{2}}$Net   & \xmark   & \checkmark    & \xmark & 82.80 & 17.42    \\
     
    MSA$^{\text{2}}$Net   & \xmark    & \xmark   & \checkmark & 83.16 & 16.18    \\
    
    MSA$^{\text{2}}$Net   & \xmark     & \checkmark   & \checkmark & 83.76 & 14.42    \\
    
    MSA$^{\text{2}}$Net   & \checkmark & \checkmark    & \xmark & 84.32 & 15.14   \\
    
    MSA$^{\text{2}}$Net   & \checkmark & \xmark    & \checkmark & 84.41 & 14.82   \\
    
    MSA$^{\text{2}}$Net   & \checkmark & \checkmark  & \checkmark & \textbf{84.75} & \textbf{13.29}      \\ \hline
    \end{tabular}}
        \caption{
       Ablation experiments on Synapse dataset:  impact of individual contributions on segmentation performance of MSA$^{\text{2}}$Net.}
    \label{table:ablation}
\end{table}

\section{Conclusion}
We proposed a novel medical image segmentation network to alleviate the restrictions of cohort studies in local and global information processing. Our key idea is to mitigate the limitations of fixed receptive sizes and basic fusion techniques in skip connections, thereby enhancing medical image segmentation, which is challenged by the variability in sizes and shapes of anatomical structures. We achieved this through a novel MASAG, which includes multi-scale fusion, spatial selection, interaction, cross-modulation, and recalibration. Our MSA$^\text{2}$Net shows robust performance across two benchmark datasets compared to SOTA methods. Ablation studies confirm the effectiveness of the design and underlying assumptions of our approach.
\section*{Acknowledgment}
The authors gratefully acknowledge the computational and data resources provided by the Leibniz Supercomputing Centre (\href{www.lrz.de}{www.lrz.de}).

\bibliography{egbib}

\clearpage

\section*{Supplementary Material}
\renewcommand{\thesubsection}{\Alph{subsection}}

\maketitle

\renewcommand{\thesubsection}{\Alph{subsection}}

\label{sec:supp}

In the supplementary material section, we provide comprehensive details about the MSA$^\text{2}$Net parameters and offer insights into the frequency analysis of the proposed MASAG module. Moreover, we visualize the attention map of the output from our decoder layers using the GradCAM \cite{selvaraju2017grad} technique to further demonstrate the effectiveness and contribution of the proposed modules. This is achieved by utilizing dynamic receptive field adjustment to capture long-range dependencies and local contextual relations among pixels, as demonstrated through GradCAM and spectral analysis.

\subsection{Frequency Analysis of MSA$^\text{2}$Net}

\begin{figure}[!ht]
    \renewcommand\thefigure{A}
    \centering \label{fig:msa2net-freq}
    \includegraphics[width=1\textwidth]{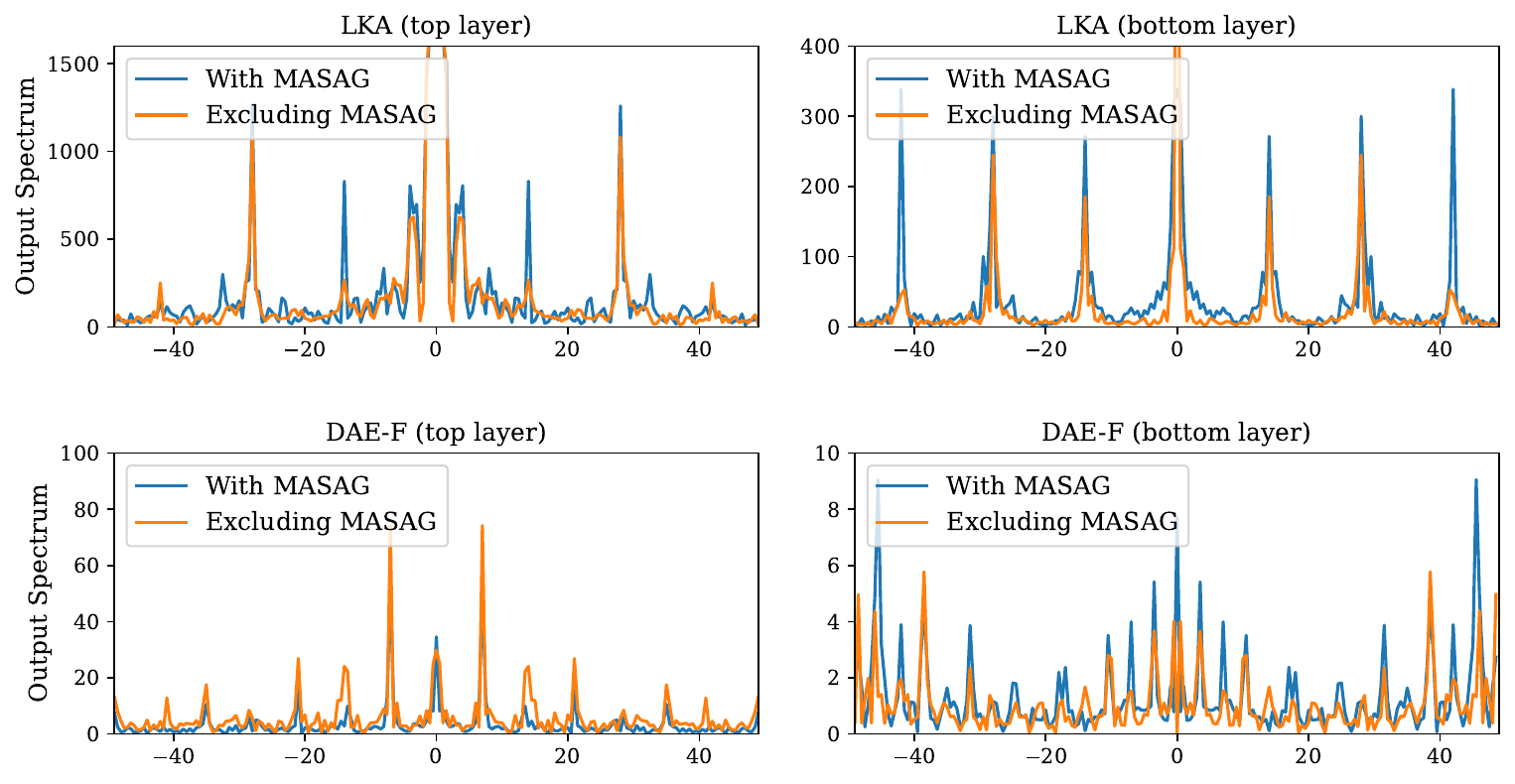}
    \caption{Frequency response analysis on the MSA$^\text{2}$Net
    (with MASAG module present) vs. MSA$^\text{2}$Net
    (excluding MASAG module).}
\end{figure}

Wang et al. \cite{wang2022anti} investigated a significant limitation of the self-attention mechanism. Through theoretical analysis, they demonstrated that self-attention functions as a low-pass filter, which removes high-frequency information and results in a loss of feature expressiveness in the model’s deep layers. The authors discovered that the Softmax operation causes self-attention to retain low-frequency information while losing fine details.
In Figure \ref{fig:msa2net-freq}, we visualize the frequency response analysis of our method with the MASAG module compared to MSA$^\text{2}$Net without it. It is evident that the standard design's frequency response in the deep layers attenuates more than that with the presence of MASAG module. This visually demonstrates the model's superior ability to preserve high-frequency details advantaging the MASAG module which can enhance high frequencies across all network layers and suppress irrelevant noise. Moreover, as shown in Figure \ref{fig:msa2net-freq} (bottom row, first column), the initial shallow layers of the decoder are prone to high-frequency details (owing to the fully transformer-based encoder), however, the incorporation of DAE-Former, LKA block, and the MASAG module aim in preserving high-frequency details, potentially affective in medical image segmentation. It should also be noted that high-frequency information is highly effective for capturing fine-grained details, especially when dealing with lesions and abnormalities that manifest as textures rather than shapes. Therefore, our MASAG module aids the network in preserving high-frequency information, which is crucial for detecting these irregular patterns.
\subsection{Analysis of Parameters}

In this section, we examine the exact number of parameters in various modules of MSA$^\text{2}$Net. Our model consists of the MaxViT-B pretrained network utilized in the encoder section, while the skip connection section incorporates three MASAG modules, and the decoder structure comprises two DAE-Former modules in the lower layers and two LKA modules in the upper layers.
Acoordingly, the encoder section (MaxViT-B) has the highest number of parameters (Table \textcolor{red}{A}). Following this, the DAE-Former section, which is entirely transformer-based, has the second-highest number of parameters. The LKA module, which employs depth-wise convolution, has fewer parameters. Finally, the total number of parameters for the three MASAG modules is 9.44 million, making it an efficient structure in terms of parameter count. Notably, our main novel module, MASAG, is highlighted
as a lightweight component, contributing only 8\% of the parameters. Owing to its low parameter count, MASAG can play a crucial role as a
plug-and-play module designed to enhance the performance of skip connections in various
networks. This efficiency makes it an attractive option for integrating into other architectures
to boost overall performance.

\begin{table}[!tbh]
        \renewcommand\thetable{A}
	\centering
	\begin{tabular}{c|c|c}
		\toprule
		\textbf{Structure}& \textbf{Module}& \textbf{Parameters (M)}\\
		\midrule
		
		Encoder & MaxViT-B &  64.12\\
		\midrule
		Skip Connection & MASAG & 9.44 \\
		\midrule
		\multirow{2}{*}{Decoder} & LKA & 1.37 \\
		\cline{2-3}
		& DAE-Former & 37.84 \\
		\midrule
		\multicolumn{2}{c}{Total}\vline & 112.77 \\
		\bottomrule 
	\end{tabular}
	\label{tab:params2}\vspace{0.5em}
        \caption{Details about the number of parameters in our model}
\end{table}

\subsection{Receptive Fields Analysis via GradCAM}
\begin{figure}[!htb]
    \renewcommand\thefigure{B}
    \centering
    \includegraphics[width=0.495\textwidth]
    {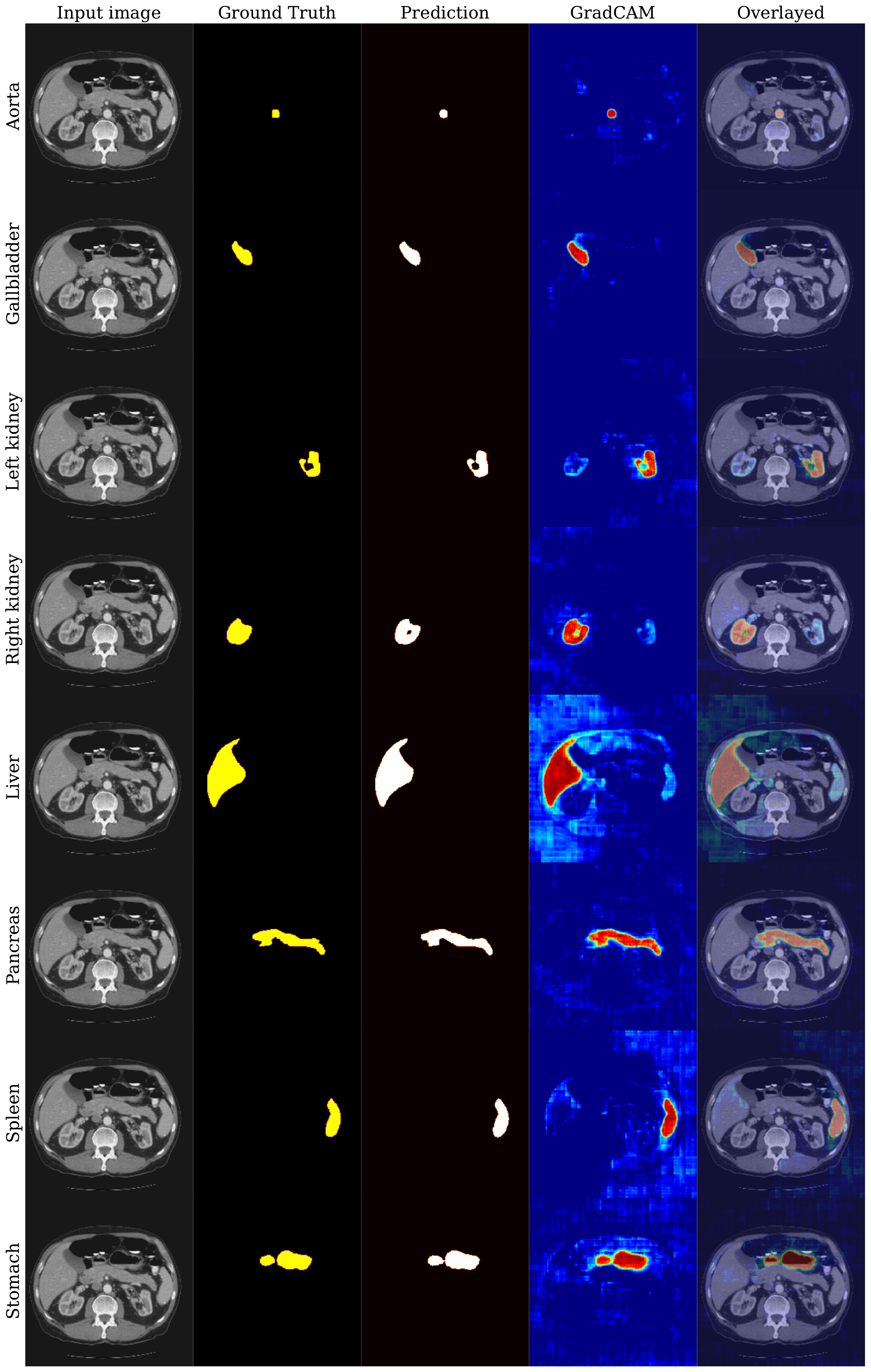}
    \hfill
    \includegraphics[width=0.495\textwidth]
    {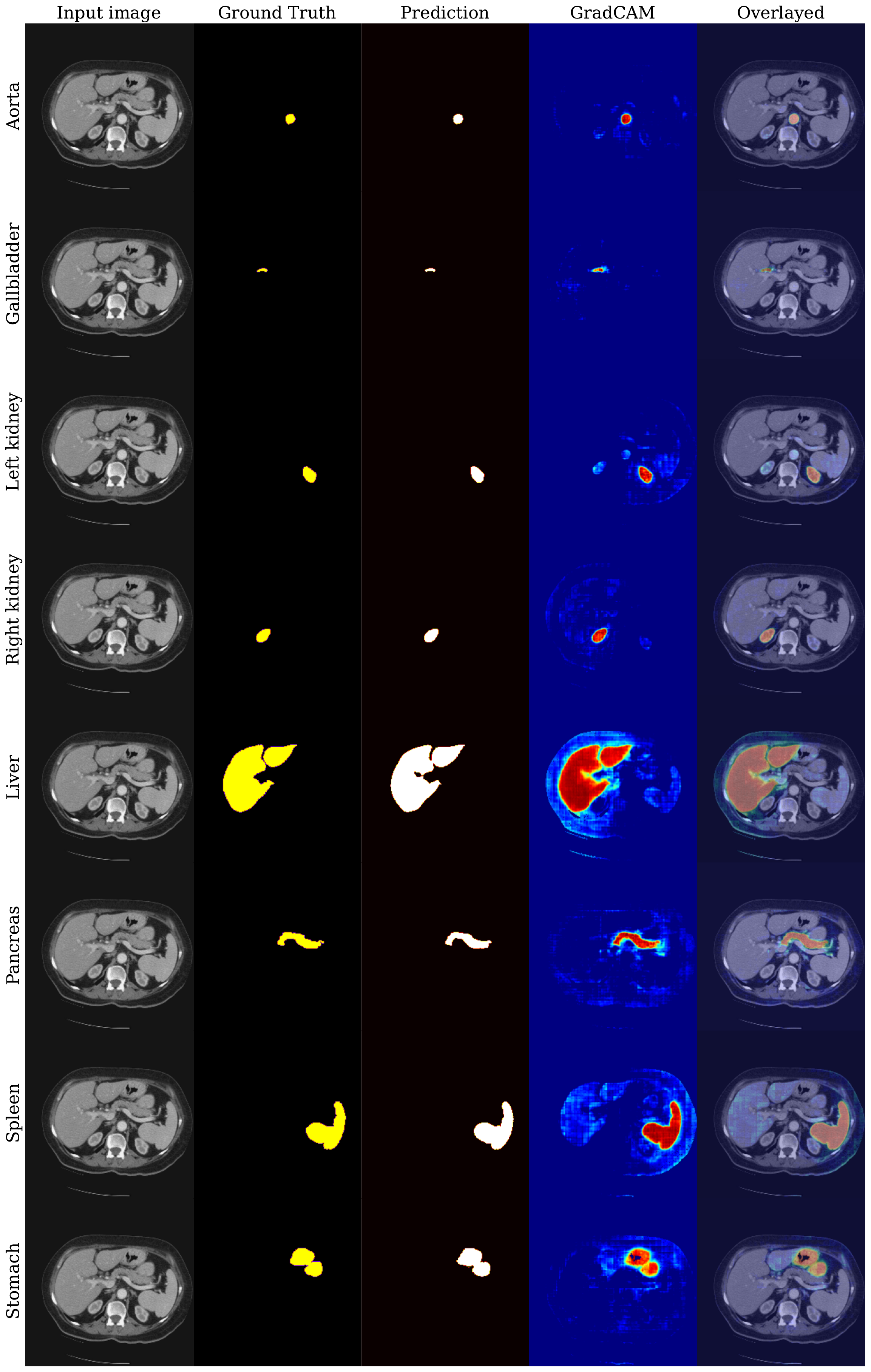}
    \caption{Feature visualization comprising two samples from the last layer of our model with different organs of the Synapse dataset (first 4 rows with smaller organs and the other larger ones). The results exhibit that with the help of the MASAG module, MSA$^\text{2}$Net is robust to the shape, size, and density variations of organs learned at the decoder.}
    \label{fig:msa2net-gradcam}
\end{figure}

The performance of the MSA$^\text{2}$Net model in detecting organs within the Synapse dataset was assessed by visualizing its attention map with GradCAM (Figure \ref{fig:msa2net-gradcam}). The results indicate that the MSA$^\text{2}$Net model is proficient at identifying small organs like the gallbladder and aorta, highlighting its ability to learn local features. Moreover, the model successfully detects larger organs, such as the liver and stomach, showcasing its capacity to capture long-range dependencies. In fact, the hybrid strategy employed in our model, incorporating the MASAG module, enables the simultaneous capture of both local and global patterns which stems from its ability in dynamic receptive field adjustment. This efficient approach significantly enhances segmentation performance by preserving high-frequency information essential for detecting fine-grained details and irregular textures, while also maintaining the ability to understand broader spatial relationships. This comprehensive capability ensures more accurate and reliable segmentation results across various organ types.

\end{document}